%% file: ajp.tex
\documentclass[12pt,aps,prb,preprint]{revtex4}

\usepackage[latin1]{inputenc}
\usepackage{color}
\usepackage{hyperref}
\usepackage{amsmath,amssymb,amsfonts,amsthm}
\usepackage{epsfig}
\usepackage{color}
\usepackage{xspace}
\usepackage{hyperref}

\usepackage{my-style}

\newtheorem{theorem}{Theorem}
\newtheorem{corollary}{Corollary}
\newtheorem{defn}{Definition}

\begin{document}

\title{No-go Theorem of Quantum Bit Commitment: Reinterpretation and Extension}

\author{Minh-Dung Dang}

\author{Patrick Bellot}
\affiliation{Institut TELECOM - TELECOM ParisTech\\
(Ecole nationale supérieure des télécommunications - ENST),\\
Department of Network and Computer Science,\\
46 rue Barrault, 75013 Paris, France}

\date{\today}

\begin{abstract}
In this article, we are interested in the physical model of general
quantum protocols implementing secure two-party computations in the
light of Mayers' and Lo's \& Chau's no-go theorems of bit commitment
and oblivious transfer. In contrast to the commonly adobted quantum
pure two-party model in the literature where classical communication
is normally ignored, we propose an alternative interpretation for
the purification of classical communication in two-party protocols by
introducing  a quantum third party for the classical channel. This
interpretation leads to a global three-party model, involving Alice's
and Bob's machines and the environment coupled to the macroscopic
channel, using the decoherence scheme in quantum measurements. This
model could give a more general view on the concealing/binding
trade-off of quantum bit commitment protocols.

Inspired from this three-party interpretation, we extend the no-go
theorems for denying some classes of two-party protocols having
access to some particular quantum trusted third-parties, known as quantum
two-party oracles. The extension implies
that a quantum protocol for implementing secure two-party computations
musts have access to a trusted third-party which erases information
and thus makes dissipation of heat to the environment.
\end{abstract}

\pacs{03.67.Dd, 03.67.Hk}
\keywords{Quantum cryptography, no-go theorem, quantum bit commitment,
quantum secure computation, trusted third-party, information erasure}

\maketitle

\section{Introduction}
Bit commitment (BC) and oblivious transfer (OT) are two fundamental
primitives of Modern Cryptography, used for the construction of secure
computations for generic two-party functions~\cite{Gol04book}. Let's
recall the definitions:
\begin{defn}[Bit Commitment]\label{defn:bc}
Bit commitment is a two-phase protocol. In the first phase,
Alice sends commitment information about a secret bit to Bob such that
Bob cannot discover Alice's secret bit from the commitment
information (\emph{concealing}). After an arbitrarily long time,
in the second phase, Alice is supposed to open the secret bit to Bob
who can successfully detect if dishonest Alice tries to open the
opposite value of the committed bit (\emph{binding}).
\end{defn}
\begin{defn}[Oblivious Transfer]\label{defn:ot}
Oblivious transfer is an asymmetrical transmission protocol
permitting Alice to send two secret bits to Bob who is allowed to
choose to get one and only one of these bits while Alice cannot know
Bob's choice.
\end{defn}

Besides, there is another closely related primitive for secure
two-party computations, named coin flipping (or coin
tossing)~\cite{Gol04book}:
\begin{defn}[Coin Flipping]\label{defn:cf}
Coin flipping is a protocol for Alice and Bob sharing a fairly random
bit, i.e. none of the parties can affect the probability distribution
of the outcome.
\end{defn}

BC and OT are equivalent as it was shown that bit commitment can be
implemented upon oblivious transfer~\cite{Kil88}, while oblivious
transfer can be built from bit commitment by transmitting quantum
information~\cite{Cre94,Yao95}. Coin flipping can be trivially
implemented upon bit commitment and oblivious
transfer~\cite{Kil88}.

Within the classical concepts of information processing, such two-party
protocols can be implemented with computational security based on
unproven assumptions of intractability in the modern computing
model~\cite{Kil88,Gol04book}. Furthermore, due to the symmetry in
trivial communications, these protocols cannot be made from scratch
with unconditional security~\cite{Kil88,Mor05dis}, as defined
by information theory~\cite{sha49}.

When passing to quantum information era, researchers have much
interest to build unconditionally secure cryptographic applications
based on special features of quantum mechanics~\cite{Wie83,BB84}. However,
while quantum key distribution had been proved
to be secure~\cite{LC99,SP00}, two no-go theorems were issued:
quantum bit commitment is impossible~\cite{May97,LC97};
quantum secure two-party computations and so oblivious transfer are
impossible~\cite{Lo97}.  Note that coin flipping is also banned
from being implemented in the scope of quantum mechanics~\cite{LC98},
even with arbitrarily small bias~\cite{Kit02}. Besides, it has been
figured out in~\cite{Ken99} that quantum bit commitment cannot be
built from coin flipping.

Mayers' and Lo's-Chau's proofs for the impossibility of quantum bit
commitment is derived from a property of pure  quantum two-party
states. In this pure quantum two-party model, a quantum bit commitment
protocol cannot be concealing and binding: if a bit commitment
protocol is secure against Bob, then Alice has a local transformation
to successfully change her secret at the opening
phase~\cite{May97,LC97}. The same property of pure quantum two-party
models was also found to deny quantum oblivious transfer in a more
sophisticated proof~\cite{Lo97}. Because of their similarity, one used
to talk only about the no-go theorem for bit commitment.

However,  the claim of the generality of the theorem caused
controversial discussions. Many doubt the validity of the model used
in the proofs for all possible hybrid protocols which could
incorporate classical computations and
communications~\cite{Yue00,Bub01,Yue04,Che03,Che05,Che06,AKS+06}. It
requires further interpretation to fit general two-party protocols
into the pure quantum two-party model.

In the literature, the no-go theorem was commonly approached in
an indirect manner using the reduction scheme. It gave only a
physical interpretation of quantum purification made by Alice and Bob
for classical computations and private random variables. The
communication of classical messages was \emph{logically} interpreted
in a reduced two-party quantum model~\cite{LC97,GL00}, not
\emph{physically}. A direct interpretation was made by Mayers who
treated the measurements made for producing classical
messages~\cite{May97}. In Mayers' interpretation, the protocol is
projected into collapsed sub-protocols corresponding to exchanged
classical messages, and the no-go theorem is applied to each pure
two-party model of each sub-protocol.

One could say that the theorems on the \emph{impossibility} of
unconditionally secure quantum bit commitment~\cite{LC97,May97}, and
on the \emph{possibility} of unconditionally secure
quantum key distribution~\cite{LC99,SP00}, are among the most
interesting subjects in the field of quantum cryptography, and
furthermore lead to philosophical feedbacks to quantum
theory~\cite{CBH03,BF05}.

\subsubsection*{Motivation and Contributions}

In this paper, we are interested in the global physical model 
interpreting all participating quantum systems in general quantum
protocols for implementing secure two-party computations, in the light
of the no-go theorems.

First, we revisit the physical model of general quantum two-party
protocols. We propose an alternative interpretation for the purification
of classical communication by introducing a quantum third party for a
macroscopic channel. This interpretation leads to a global three-party
model, involving Alice's and Bob's machines and the environment
coupled to the classical channel, referring to the decoherence model
in quantum measurements. We also find that, in this three-party model,
if a bit commitment protocol is concealing, than Alice has a local
transformation to cheat. This three-party model interprets the
physical systems in general protocols more faithfully than the reduce
two-party model in the literature~\cite{LC97,GL00}. Moreover, it could
give a more global view on the average concealment and binding
parameters of a general protocol, in comparison with
Mayers' one which treats each individual sub-protocol~\cite{May97}.

Inspired from this three-party interpretation, we extend the no-go
theorems for two-party protocols having access to some particular
quantum trusted third-parties, named quantum two-party oracles, where
no information is erased from the view of Alice and Bob. This
extension covers a no-go result similar to Kent's one for
coin-flipping-based protocols~\cite{Ken99}.

This extension of no-go theorems implies that a quantum protocol for
implementing secure two-party computations musts have access to a
trusted third-party which erases information and thus makes
dissipation of heat to the environment. Nevertheless, we can build
classical oracles which do logical reversible computations for
implementing oblivious transfer. That leads to a discussion on the
physical nature of classical information.

\subsubsection*{Organization of The Paper}
In Section~\ref{sec:nogo-thm}, we expose an overview on the no-go
theorem of quantum bit commitment. In
Section~\ref{sec:macro-channel}, we present our physical
interpretation for general two-party protocols with a three-party model
regarding the presence of a macroscopic channel. In
Section~\ref{sec:ext-nogo}, we do some studies  on particular
oracle based protocols which are penalized by the no-go
theorem. Then in Section~\ref{sec:cf-based}, we show that coin
flipping belong to this class of oracles and cannot help to implement
bit commitment. Finally, in Section~\ref{sec:thermodynamics}, we
discuss the implementation of unconditionally secure two-party
computations' primitives and the physic of classical information from
a thermodynamical view point.

\section{Overview on No-go Theorem of Quantum BC}\label{sec:nogo-thm}
\subsection{Canonical Theorem}\label{sec:canonical-nogo}
In this section, we expose the canonical no-go theorem for quantum
deterministic protocol executed on a pair of Alice and Bob quantum
machines which interact by communicating quantum
signal.
Formally, a computation is an evolution in time of the
\emph{computational configuration} that consists of
variables (systems) which are assigned with values (states) following
a prescribed algorithm.
In the quantum world the configuration at one moment is
described by the state of all participating quantum systems at that
moment.  For a two-party protocol, the transition from one configuration to
another successive configuration is made by local unitary transformations at
Alice's and Bob's sides and by the communications between them.
The communication of quantum signal is considered as quantum
particles are faithfully brought from sender's machine to receiver's
machine.

Any bit commitment protocol can be seen as a two-phase computation,
jointly made by Alice and Bob. After the first phase - commit phase, the
computation is interrupted, and then continued in the second phase -
opening phase. The computation takes Alice secret bit to be committed
to Bob as input, and give one of three outputs: $0$ - if Bob is
convinced that Alice's input is $b=0$; $1$ - if Bob is convinced that
Alice's input is $b=1$; and $\perp$ if any cheating user is detected
by the other.

As the detection of Bob's cheating would rather be made before the
opening phase, we are only interested in the privacy against Bob's
(concealment) and the detection of Alice's cheating (binding), once
the commit phase has ended, i.e. the computation has been interrupted.

In a deterministic bit commitment protocol, according to a
deterministic algorithm, Alice and Bob must prepare two quantum systems
$A$ and $B$, characterized by $\hilbert{} = \hilbert{A,init} \otimes
\hilbert{B,init}$, initially in a certain determined pure state
$\ket{\psi(b)_{init}} = \ket{\psi(b)}_{A,init}\ket{0}_{B,init}$. At
step $i$, Alice and Bob realize a joint
computation $U_i = U_{A,i} \otimes U_{B,i}$ on $\ket{\psi(b)_{i-1}}$
to get $\ket{\psi(b)_i}$
and exchange some subsystems in communication. Then, the configuration
$\ket{\psi(b)_i}$ is split into two parts according to the new
decomposition $\hilbert{} = \hilbert{A,i} \otimes
\hilbert{B,i}$. Here, $\hilbert{}$ is invariant, but its decomposition
into Alice and Bob's parts varies with communications.  For
simplifying, we will use $\hilbert{A}, \hilbert{B}$ instead of
$\hilbert{A,i}, \hilbert{B,i}$ to implicitly  specify the
decomposition at the moment of speaking.

The computation is then a determined sequence of configurations
$\ket{\Psi(b)_{init}}, ..,\ket{\Psi(b)_{final}}$. At step $i$, the
corresponding configuration $\ket{\Psi(b)_i}$ is split into two
partial configurations at Alice and Bob sides:
\begin{align*}
\rho^A(b)_i &= tr_{B}(\projection{\Psi(b)_i}{\Psi(b)_i}),\\
\rho^B(b)_i &= tr_{A}(\projection{\Psi(b)_i}{\Psi(b)_i}).
\end{align*}
If the protocol is unconditionally concealing then Bob have not to be
able to distinguish $\rho^B(0)_i$ from $\rho^B(1)_i$  for all $i \leq
int$ where $int$ is the interruption step, i.e. $\forall i \leq int,
\rho^B(0)_i = \rho^B(1)_i$. Here, it suffices to be only interested in
$\rho^B(0)_i = \rho^B(1)_i$ at the interruption step $i = int$.

We could expect that Alice cannot replace $\rho^A(0)_{int}$ with
$\rho^A(1)_{int}$ and vice-versa because of the \emph{entanglement} in
$\ket{\Psi(b)_{int}}$. Unfortunately, following \cite{HJW93}, in case
$\rho^B(0)_{int} = \rho^B(1)_{int}$, there exists a unitary
transformation $U_A$ acting in $\hilbert{A}$ that maps
$\ket{\Psi(1)_{int}}$ into $\ket{\Psi(0)_{int}}$. Therefore, Alice can
replace the partial configuration by the operators $U_A$ and $U_A^{-1}$.

More generally, quantum model allows a non-ideal unconditional security,
i.e $\rho^B(0)_{int} \approx \rho^B(1)_{int}$.
The security of Alice's bit can be measured by the distinguishability
between $\rho^B(0)_{int}$ and $\rho^B(1)_{int}$, for instance the
fidelity of quantum states:
\begin{equation}
F(\rho^B(0),\rho^B(1)) \geq 1 - \epsilon.
\label{equa:concealing}
\end{equation}
The extension of Uhlmann's theorem \cite{HJW93}(\cite{NC04book} -
exercise 9.15) states that there exists a purification
$\ket{\Psi'(0)_{int}}$ of $\rho^B(1)_{int}$ such that
$$
|\bracket{\Psi(0)_{int}}{\Psi'(0)_{int}}| =
F(\rho^B(0)_{int},\rho^B(1)_{int}) \geq 1 - \epsilon.
$$
Recall that, as $\ket{\Psi'(0)_{int}}$ and $\ket{\Psi(1)_{int}}$ are two
purifications of $\rho^B(1)_{int}$, there exists a unitary transformation
for Alice to switch between $\ket{\Psi'(0)_{int}}$ and
$\ket{\Psi(1)_{int}}$. Therefore, suppose that Alice has began the
computation for $b = 1$, she can cheat by transforming $\ket{\Psi(1)_{int}}$
into $\ket{\Psi'(0)_{int}}$ and declaring $b=0$. The opening phase
will be continued with $\ket{\Psi'(0)_{int+1}},
...\ket{\Psi'(0)_{final}}$ under unitary transformations. So:
$$
|\bracket{\Psi(0)_{final}}{\Psi'(0)_{final}}| \geq 1 - \epsilon.
$$
A measure for the success of Alice's cheating is
$$F(\rho^B(0)_{final},\rho'^B(0)_{final}).
$$
Following Uhlmann's theorem (\cite{NC04book} - theorem 9.4), we have
\begin{align}
F(\rho(0)^B_{final},\rho'^B(0)_{final}) &\geq
|\bracket{\Psi(0)_{final}}{\Psi'(0)_{final}}| \nonumber \\
	&\geq 1 - \epsilon.\label{equa:binding}
\end{align}
Therefore, in a pure deterministic quantum model, we cannot have a bit
commitment protocol that is both concealing and binding. Moreover, the
more a protocol is concealing, the more it is binding, by the measure
of quantum fidelity,
cf. Eqs. \eqref{equa:concealing},\eqref{equa:binding}.

\subsection{Interpretation for Generality}\label{sec:interpretation}
A major objection to MLC no-go theorem is that it is ``too simple to
be true'' for all possible protocols where Alice and Bob
\begin{enumerate}
\item do measurement on their quantum systems and pass to classical
computation; and introduce secret variables;
\item communicate classical information through a macroscopic channel
that does permit to transmit quantum signal.
\end{enumerate}

Most of attention were paid to explain private classical variables in
computations~\cite{Yue00,Bub01,Yue04,Che03,Che06}. With
unbounded quantum machines, Alice and Bob are allowed to keep all of
the computations at quantum level where all probabilist choices can be
purified by appropriate additional quantum dices. For instance, to
create a classical binary random variable $x$, in state $\ket{0}$ or
$\ket{1}$ with probability $1/2$, one prepares two system $x, y$
jointly in entangled state $(\ket{0_x0_y} + \ket{1_x1_y})/\sqrt{2}$
and use only system $x$ for the computation. The purified protocol
becomes deterministic, acting on a larger quantum system encompassing Alice's
and Bob's additional dices for purifying private classical random
variables. Then, as shown in the canonical theorem, this two-party
purified protocol cannot implement bit commitment.

In \cite{Yue02,Yue04}, Yuen raised the problem of secret variables which
questioned that if Bob really did the measurements, then the whole
system might be projected into a secret collapsed state corresponding
to Bob's secret value and Alice would not know the corresponding cheating
transformation. This was successfully treated in \cite{Bub01,Che06},
showing that Alice's cheating transformation is universal for ideal and
nearly ideal protocols.

However, the classical communication is normally omitted with
some assumptions on the communication: ``classical communication can
be carried out by quantum model, but with some
constraints''~\cite{LC97}. But what are the constraints?

Imagine that in the specification of a protocol, at a certain
moment, a party $S$ has to measure a certain quantum state $\ket{\psi}_S$
with an apparatus of $n$ degrees of freedom and communicate the
outcome to the other via a classical channel. This measurement will
output $i \in \{1,..,n\}$ with probability $p(i)$ and set the measured
system in state $\ket{\psi_i}_S$. Receiving the classical value $i$, the
receiver's apparatus $R$ generates $n$-dimension variable in basis state $\ket{i}_R$ for further computations.

Of course, we can reduce this communication to a pure two-party
quantum model where the sender realizes a transformation
$$
U(\ket{\psi}_S \otimes \ket{0}_R) \rightarrow \sum_{i=1}^n \sqrt{p(i)}
\ket{\psi_i}_S\otimes\ket{i}_R
$$
and the protocol is emulated correctly because the density-matrix
description of each system is the same as though a real measurement
is done~\cite{LC97,Bub01}. The protocol configuration is then reduced
to a two-party model consisting of Alice's and Bob's machines. And by
the purification of private random variables, the reduced model
becomes a pure quantum two-party state which is not allowed to
implement bit commitment.

However, the above reduced model for classical communications does not
interpret what really happen in the physical world.
From the physical view point, the classical channel does not appear in
this reduced two-party quantum model.
Indeed,  in a generic protocol, the communication of classical
messages forces destroying the purity of two-party states.

What is the difference between a quantum channel and a classical one?
A quantum channel is a medium that we can use to directly transmit a
quantum state without disturbing it. Nevertheless a classical channel,
for transmitting discrete messages, permits only one from a collection of
discrete signal values which can be amplified by many \emph{quantum
systems} on the channel, for instance a macroscopic electrical wire
with tension $+5V$ for $0$ and $-5V$ for $1$. So, a classical channel
forces the measurements to be done for
making classical signals i.e. Alice and Bob have to really measure
their quantum states to make classical messages. The real joint
computation with communication by measuring and transmitting classical
values via a classical channel is not an evolution of a pure two-party
state.  In other words, as the action of measurements ``can never help a
cheater'' \cite{GL00}, why it does not prevent Alice from cheating?

This point was only explained in Mayers'
version where the measurements for making classical messages were
considered~\cite{May97}. Following Mayers, Alice and Bob would keep
all of the operations at the quantum level,
except for making classical messages. Thus, for each classical message
$\gamma$, the corresponding quantum system is collapsed to a known
pure two-party state $\ket{\psi_{b,\gamma}}_{AB}$, and the trade-off
between the concealment and the binding is separately treated for this
state, i.e. if the collapsed protocol conceals:
\begin{align}
F_\gamma &= F\left(\rho_\gamma^B(0), \rho_\gamma^B(1)\right) \nonumber \\
	&= F(tr_A(\projection{\psi_{0,\gamma}}{\psi_{0,\gamma}}),
tr_A(\projection{\psi_{1,\gamma}}{\psi_{1,\gamma}}))\nonumber\\
	&\geq 1 - \epsilon \label{equa:mayers-concealment}
\end{align}
then Alice has a unitary cheating transformation $U_{A,\gamma}$ with
possibility of success
\begin{equation}
|\bracket{\psi_{0,\gamma}}{U_{A,\gamma}|\psi_{1,\gamma}}| = F_\gamma
\geq 1 - \epsilon.\label{equa:mayers-binding}
\end{equation}

\section{Three-party Model for a Macroscopic
Classical Channel}\label{sec:macro-channel}
We suppose that Alice and Bob implement a two-party protocol,
communicating quantum signal via a quantum channel and classical
signal via a macroscopic channel. As analyzed in
Section~\ref{sec:canonical-nogo}, the communications of quantum
messages make only repartitions of quantum computation systems in
Alice and Bob's machines. We also suppose that Alice and Bob have
unlimited quantum machines for purifying all private classical random
variables. Nevertheless, the measurements for making classical
messages to be exchanged via the macroscopic channel cannot be
purified by Alice's and Bob's dices.


It is natural to interpret that in reality a classical channel is 
coupled with the environment where the decoherence is so strong that the
messages transmitted on the channel are measured by a CNOT-like gate,
copied, and amplified by an infinite quantum systems in the
environment, i.e. a basis qubit $\ket{i}$ becomes
$\ket{i}\otimes\ket{i}_E$~\cite{Zur91,BS98}.

Suppose that the process of communication of classical message via a
classical channel as follows:
\begin{enumerate}
\item The sender $S \in \{A,B\}$ has to measure some quantum state
$\ket{\psi}_{AB}$ with an apparatus with $n$ degrees. This measurement
will output $i \in \{1,..,n\}$ with probability $p(i)$ and let the
measured system in a state $\ket{\psi_i}_{AB}$:
$$
\ket{\psi} \rightarrow \sum_i \sqrt{p(i)}
\ket{\psi_i}_{AB}\ket{i}_S\ket{i}_{E,S}
$$
where $\hilbert{E,S}$ is for the macroscopic part in the measurement
device lost to the environment that causes the impurity of
sender's state.
\item The sender sends the signal $i$ via a macroscopic channel where
the signal can be infinitely amplified by the environment $E$:
$$
\ket{i}_S \rightarrow \ket{i}_S\otimes\ket{i}_E.
$$
\item The signal is amplified, and propagates to the receiver's
device, where the corresponding quantum
state $\ket{i}$ will be generated for the receiver's quantum machine
$R = \{A,B\} \setminus \{S\}$:
$$
\ket{i}_E \rightarrow \ket{i}_E \otimes \ket{i}_R.
$$
\end{enumerate}

Therefore, this process is a unitary transformation acting on a pure
state, but in a larger space covering Alice's, Bob's machine and the
environmental systems amplifying the signals:
$$
\ket{\psi}_{AB}\ket{0}_{S,R,E*} \rightarrow \sum_{i=1}^n
\sqrt{p(i)}\ket{i}_S\ket{i}_{E*}\ket{i}_R\ket{\psi_i}_{AB}
$$
where $E*$ denotes all systems of the environment, and $S,R$ denote
the controllable quantum systems in Alice's and Bob's machines. The
initial states of systems amplifying classical messages process are
not important, and denoted by $\ket{0}_{S,R,E*}$. So, by
introducing the environment systems $E*$, the execution of the
protocol is seen as a deterministic unitary evolution of the global
three-party state lying in
$\hilbert{A}\otimes\hilbert{B}\otimes\hilbert{E*}$.

Here, $\hilbert{E*}$ is not controlled by any participant, and the
configurations of the protocol are not pure states lying in a
two-party space for quantum systems in Alice' and Bob's machines
anymore. In fact, it is a three-party model where the systems in
$E*$ play a passive role via the CNOT gates.

Therefore, the protocol is a deterministic computation in a
three-party space and the configuration of the protocol at any moment
can be described by a known pure state in the form of
\begin{equation}
\ket{\Psi(b)} = \sum_{i=1}^N \sqrt{p_b(i)} \ket{i}_{E*}\ket{i}_A
\ket{i}_B \ket{\psi_i(b)}_{AB}
\label{equa:purified-classical-messages}
\end{equation}
where $i$ is any possible classical message, and
$\ket{i}_A,\ket{i}_B$ appear for the fact that Alice and Bob
should duplicate and keep a record of the classical messages
forever in their machines.


For the security on Bob's side, the protocol has to assume
\begin{equation}
F(\rho^B(0),\rho^B(1)) \geq 1 - \epsilon
\label{equa:global-conceal}
\end{equation}
where $\rho^B(b) = tr_{E*}(tr_A(\projection{\Psi(b)}{\Psi(b)}))$.

Obviously,
\begin{equation}
F(\rho^{B,E*}(0),\rho^{B,E*}(1)) \leq
F(\rho^B(0),\rho^B(1))\label{equa:noiseless-equality}
\end{equation}
where $\rho^{B,E*}(b) = tr_A(\rho(b))$, and Alice  can only control
the quantum systems in his machine $\hilbert{A}$. We would expect that
this inequality can help to prevent Alice cheating when
$F(\rho^{B,E*}(0),\rho^{B,E*}(1)) \ll 1$. However, the inequality
happens when information are lost during communication via the
classical channel.
Unfortunately, the environment has only honestly amplified the
signals and the equality is obtained:
\begin{align*}
F(\rho^{B,E*}(0),\rho^{B,E*}(1)) &= F(\rho^B(0),\rho^B(1))\\
	&\geq 1 - \epsilon
\end{align*}
because in the description of $\ket{\Psi(b)}$, $\ket{i}_{E*}$ is
exactly the same as $\ket{i}_A$. The classical channel is noiseless
and does not help. We could recall that a noisy channel could enable
us to build unconditionally secure primitives~\cite{Cre97,CMW04}.

There exists a unitary transformation $U_A$ such that
\begin{equation}
\left | \bracket{\Psi(0)}{U_A|\Psi(1)}\right | \geq 1 - \epsilon,
\label{equa:global-cheat}
\end{equation}
and Alice can use it to cheat.

The above purified model exists only if we accept the concept of
decoherence that leads to the \emph{Many Worlds Interpretation} of
quantum mechanics where the pure global state exists as the
multiverse of classical realms corresponding to the collapsed
state~\cite{Sch04}. This pure state may not exist in reality
according to the \emph{Copenhagen Interpretation}, because Alice and
Bob should be in one of $N$ situations, provided a collapsed state
$\ket{i}_A\ket{i}_B\ket{\psi_i(b)}_{AB}$ with the corresponding
probabilities $p_b(i)$, i.e. we are provided instead a statistical
ensemble $\{p_b(i), \ket{i}_A\ket{i}_B\ket{\psi_i(b)}_{AB}\}$.

In that case, Alice's average
cheating possibility over all occurrences of exchanged classical
messages can be measured by
\begin{align}
\sum_i^N \sqrt{p_0(i)p_1(i)}
|\bra{\psi_i(0)}\bra{i}U_A\ket{i}\ket{\psi_i(1)}| &\geq \left |
\bracket{\Psi(0)}{U_A|\Psi(1)}\right |\nonumber\\
	&\geq 1 - \epsilon\label{equa:average-cheat}
\end{align}

We see that, a protocol mays not necessarily conceal against Bob for
all collapsed sub-protocols but on average; and then Alice mays not
successfully cheat for all collapsed sub-protocols but on average, cf.
Eqs. \eqref{equa:global-conceal}, \eqref{equa:average-cheat}.

In comparison with Mayers' model, we see that these collapsed states are
the same as $\ket{\psi_{b,\gamma}}$ for $i =
\gamma$. cf. Eqs. \eqref{equa:mayers-concealment}
\eqref{equa:mayers-binding}. However, we could relax the requirement
for each sub-protocol, for example, $F_\gamma$ could be small for some
$\gamma$ but the occurring probability of $\gamma$ is small. Moreover,
it can happen that the occurring probabilities of $\gamma$ for the
commitment of $0$ and $1$ are different, i.e. $p_0(\gamma) \neq
p_1(\gamma)$.

The above concealment, cf. Eq. \eqref{equa:global-conceal}, suggests
to extended the average concealment for the protocol based on Mayers'
individual collapsed protocols:
$$
CONC'  = \sum_\gamma \sqrt{p_0(\gamma)p_1(\gamma)} F_\gamma
$$
If we could measure the average concealment by
\begin{align*}
CHEAT' &= \sum_\gamma \sqrt{p_0(\gamma)p_1(\gamma)}
|\bracket{\psi_{0,\gamma}}{U_{A,\gamma}|\psi_{1,\gamma}}|\\
	& = \sum_\gamma \sqrt{p_0(\gamma)p_1(\gamma)}F_\gamma
\end{align*}
then $CHEAT'= CONC'$. Moreover, as a standard, the concealment can be
measured by
$$
CONC = F\left(\sum_\gamma p_0(\gamma)\rho_\gamma^B(0), \sum_\gamma
p_1(\gamma)\rho_\gamma^B(1)\right).
$$
Normally, $CHEAT' \leq CONC$ (\cite{NC04book} - theorem 9.7), but as
Bob keeps a record of classical message $\gamma$ in his quantum
state $\rho_\gamma^B(b)$ the two measures of concealment are identical
$CONC' = CONC$ and then $CHEAT' = CONC$.

\begin{figure}[htb]
\begin{center}
\resizebox{0.4\textwidth}{!}{\input{global-purified-model.pstex_t}}
\caption{The global purified model}\label{fig:global-model}
\end{center}
\end{figure}
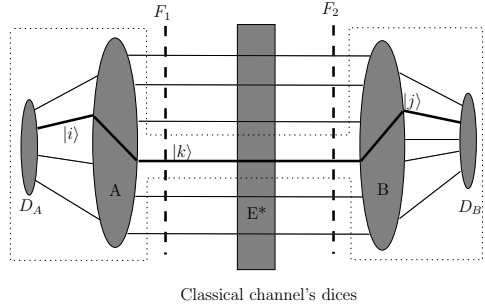

In summary, the global purified model which is obtained by the
purification of local random variables and exchanged classical
messages can be illustrated as in Fig.~\ref{fig:global-model} which
describes the configuration of the protocol at any given moment. This
configuration is in a pure state:
$$
\ket{\Psi(b)} =
\sum_{k,i,j}\sqrt{p_b(k,i,j)}\ket{k}_{ABE*}
\ket{i}_{D_A}\ket{j}_{D_B}\ket{\psi_{k,i,j}(b)}_{AB}
$$
We represent each entanglement connection via a
classical value $i, j, k$ by a line through the concerned space.
For instance, the real configuration of the
protocol corresponding to Alice's private outcome $i$, Bob's private
outcome $j$ and exchanged classical message $k$ is represented by the
bold line in the figure. The execution of the protocol is a sequence
of deterministic unitary transitions between successive
configurations. It is a parallel execution of
many honest schemes.

As Alice and Bob have the possibility to keep their dices in their
quantum machines, we would throw $D_A$ to $A$ and $D_B$ to $B$ and the
the no-go theorem is applied to the model as analyzed above.

Note that, if the purification of local variables $\ket{i}$ and
$\ket{j}$ is really possible as Alice's and Bob's throw the
private dices $D_A,D_B$ to their quantum machines, the purification of
exchanged classical messages $\ket{k}$ is more abstract. It is a
quantum parallelism of
collapsed counterparts corresponding to exchanged classical messages
as in Mayers' interpretation~\cite{May97}:
the configuration corresponding to the classical message $k$ lies in
the region marked by the dot line in the figure.

This global purification describes the real execution of
a protocol only if the Nature follows the theory of \emph{Decoherence}
and \emph{Many Worlds Interpretation}. In any way, it is a convenient
model for analyzing the average values of concealment and binding of
general protocols.

Logically, we are allowed to reduce this three-party model to a pure
quantum two-party model by making $\ket{i}_{E*}$ disappear as this is
only a redundant copy of $\ket{i}_{A}\ket{i}_{B}$.  The frontier $F_1$
at the limit of Alice control gives Bob the same information as at
$F_2$. However, this reduced pure quantum two-party model only
emulates the real protocols \emph{logically}, \emph{not
physically}. Thus, the reduction could not be evident without a
physical interpretation.

\section{Trusted Third-Party penalized by No-go
Theorems}\label{sec:ext-nogo}
Because of the no-go theorem on two-party protocols, we could be
satisfied to use a trusted third party for unconditionally
secure computations. It is trivial when we have a trusted third party
for implementing these protocols. For instance, in an oblivious
transfer protocol, Alice sends $b_0,b_1$ and Bob sends $c$ to Trent
who is honest; Trent sends $b_c$ to Bob.

In a general case, we construct a trusted two-party circuit for any
desired computation, with some inputs from Alice and Bob, and some
outputs back to Alice and Bob.  The execution time of the computation
done by the oracle is an elementary unit, and  the results are
immediately returned to the users. We name such trusted third-party as
\emph{two-party oracle} model.

In this section, present an extension of the impossibility of quantum
bit commitment and oblivious transfer for some common quantum two-party
oracle models.

\subsection{Quantum Resource-Limited Oracle}
In the actual situation, we may have a quantum oracle but with limited
resources. A common kind of quantum oracles is the quantum-circuit
model where the oracles are built as quantum circuits for the required
function without any resources. This model of quantum oracles is much
used in quantum computation~\cite{NC04book}.

However, such a quantum oracle would have to throw information
acquired by measurements during the computation into another public
environment to re-initiate its private state for further usage. Thus,
the configuration of any two-party protocol using this oracle is not
correlated with the private resources of the oracle. So, such a
short-term oracle cannot help to build secure two-party computations,
for long-term usage.

\subsection{Quantum Post-Empty Oracle}\label{sec:pe-oracle}
We define here a class of quantum oracles as quantum circuits with
some private resources for the inputs but sending all of the outputs
to the users.

\begin{defn}\label{defn:pe-oracle}
A \emph{Post-Empty Oracle} (PE-O) is defined as a two-party oracle that
implements any specified algorithm, using some local variables. In the
end of the computation, the oracle splits all of the final variables,
including the local ones, and sends back one part to Alice, one part
to Bob.
\end{defn}

\begin{figure}[htb]
\begin{center}
\resizebox{0.4\textwidth}{!}{\epsfig{file=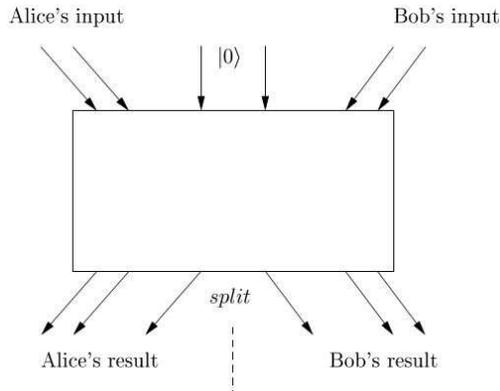}}
\caption{Quantum Post-Empty Oracle}\label{fig:trusted-computation}
\end{center}
\end{figure}

Figure~\ref{fig:trusted-computation} illustrates a quantum PE-O: it
receives quantum signal for inputs from Alice and Bob; initializes
necessary local variables to $\ket{0}$; applied the required
computation to these inputs; and at the end splits all of the outputs,
including the local variables, into two parts, redirects one part to
Alice, and one part to Bob.

We can extend the no-go theorems to more general quantum two-party
protocols having access to PE-Os:

\begin{theorem}\label{thm:pe-oracle}
We cannot build secure quantum bit commitment and oblivious transfer
protocols with quantum PE-Os.
\end{theorem}

It is then sufficient to prove the theorem for the deterministic
purified reduced model. The interpretation for general protocols with
classical communications has been analyzed in
Section~\ref{sec:macro-channel}.

\begin{proof} (Sketch).
In fact, when the oracle splits all of quantum input and local
variables which participate to the computations to Alice and Bob, the
global configuration is in some known pure state, according to the
algorithm, in a two-party space. Thus, such a model cannot implement
bit commitment and oblivious transfer.

Indeed, for the commitment of $b$, Alice, Bob and the oracle must
prepare three quantum systems $A$, $B$ and $T$, characterized by
$\hilbert{} = \hilbert{A} \otimes \hilbert{B} \otimes \hilbert{T}$,
initially in some determined pure state
$$
\ket{\Psi(b)} =
\ket{\psi(b)}_{AB}\otimes\ket{0}_{T_1}...\ket{0}_{T_n}
$$
where $n$ is the number of requests that Alice and Bob appeal to the
oracle during the protocol, and $T_i$ is the local quantum variable
used in $i^{th}$ call.

At any step, after $i$ calls to the oracle, the configuration of the
protocol is
$$
\ket{\Psi_i(b)} =
U(\ket{\psi(b)}_{AB}\ket{0}_{{AB}_1}...\ket{0}_{{AB}_i}) \otimes
\ket{0}_{T_{i+1}}...\ket{0}_{T_n},
$$
for a certain $U$ where the system $T_i$ has been split into
$AB_i$. The corresponding partial configuration at Bob side will be:
$$
\rho^B(b)_i = tr_{A,T}(\projection{\Psi(b)}{\Psi(b)}).
$$
Then, we see that after the commitment phase, at a certain step $c$,
when $\rho_c^B(0) = \rho_c^B(1)$, there exists a unitary
transformation $U_A$ acting in $\hilbert{A}$ that maps $\ket{\Psi_c(1)}$
into $\ket{\Psi_c(0)}$. Therefore, Alice can cheat by switching the
partial configuration with the operators $U_A$ and $U_A^{-1}$.
\end{proof}

\subsection{Quantum Trivial Oracle}
A trivial case is that we may have an oracle with unlimited resources,
but it could not hide information from Alice and Bob.

\begin{defn}\label{defn:trivial-oracle}
A Quantum Trivial Oracle is defined as a two-party oracle
which implements the computation of any two-party function. The oracle
can be coupled with a quantum system $O$. But whenever the oracle
acquires information into its memory $O$ by measurements, the
information is thrown into the public environment and observed by
Alice and Bob.
\end{defn}

Then, more generally, we can extend the no-go theorems to quantum
protocols based on such trivial oracles.

\begin{theorem}\label{thm:trivial-oracle}
We cannot build secure quantum bit commitment and oblivious transfer
protocols based on Quantum Trivial Oracles.
\end{theorem}

\begin{proof} (Sketch).
Let's return to the global model purifying classical messages of a
quantum two-party protocol,
cf. Eq. \eqref{equa:purified-classical-messages}. In a protocol having
access to quantum trivial oracles with a supplementary system $O$, we
can throw all of systems in $O$ to the global third party environment
$E*$. As the oracles send copies of the information thrown to $O$ to
Alice and Bob, the global configuration at any moment of a bit
commitment protocol is in the same form of
$$
\ket{\Psi(b)} = \sum_{i=1}^N \sqrt{p_b(i)} \ket{i}_{E*}\ket{i}_A
\ket{i}_B \ket{\psi_i(b)}_{AB}
$$
and if the protocol does conceal then it cannot be binding.

In summary, in this three-party model involving Alice's machine, Bob's
machine and the systems in $E*$:
\begin{itemize}
\item The systems in $E*$ do not hide information from Bob in a bit
commitment scheme. The global model can be considered as a two-party
model $\hilbert{A} \otimes (\hilbert{E*} \otimes \hilbert{B})$ where
$\hilbert{E*} \otimes \hilbert{B}$ is for what Bob can learn about
Alice's secret and $\hilbert{A}$ is for what Alice can fully control
to cheat.
\item The systems in $E*$ do not hide information from Alice in an
oblivious transfer scheme. The global model can be considered as a
two-party model $(\hilbert{A} \otimes \hilbert{E*}) \otimes
\hilbert{B}$ where $\hilbert{A} \otimes \hilbert{E*}$ is for what
Alice can learn about Bob's secret and $\hilbert{B}$ is for what Alice
can fully control to cheat.
\end{itemize}
\end{proof}
In our interpretation in Section~\ref{sec:macro-channel}, the
macroscopic channel
for Alice and Bob communicating classical information plays the role
of a trusted oracle. But this oracle is trivial as it publicly
measures the quantum systems of Alice and Bob machines, and the
measurement results are observed by Alice and Bob. The measurements
for making classical messages are not information-erasing in the joint
view of Alice and Bob.

\section{Coin-Flipping based Protocols}\label{sec:cf-based}
\begin{corollary}
Coin Flipping based Quantum Bit Commitment and Quantum Oblivious
Transfer are impossible.
\end{corollary}

In \cite{Ken99}, Kent shown a similar result. In his paper, he
established a relativist model to implement coin flipping. With the
model of  quantum two-party oracle, we make the statement more
comprehensible from a non-relativist point of view.

\begin{proof}
We can state that coin flipping is weaker than
bit commitment and oblivious transfer in a reduction style. Indeed, we
suppose that Alice
and Bob have access to a PE-O that creates a
pair of qubits in Bell state $\ket{\Phi+} = (\ket{0}_A\ket{0}_B +
\ket{1}_A\ket{1}_B)/\sqrt{2}$ and sends each part to a user. With
such a PE-O, Alice and Bob has a fair quantum coin that
can realize classical coin flipping: Alice and Bob measure
$\ket{\Phi+}$ in the same basis $\{\ket{0},\ket{1}\}$ to share a
random bit. However, quantum  bit commitment and oblivious transfer
are not realizable with this PE-O, as shown by Theorem
\ref{thm:pe-oracle}.


Besides, we show here a more direct proof for protocols based on classical
coin flipping. Suppose that Alice and Bob have access to an oracle
that generates classical random coins and send two copies to Alice
and Bob. In fact, the pair of classical coins is a probabilistic
ensemble of $\ket{0}_A\ket{0}_B, \ket{1}_A\ket{1}_B$
with probabilities $1/2, 1/2$:
$$
\rho^{AB} = (\projection{0_A0_B}{0_A0_B} + \projection{1_A1_B}{1_A1_B})/2
$$
These coins should be represented by a pure state in an augmented model
as though they are entangled with a third-party system $O$.
$$
\ket{C} = \sqrt{1/2}(\ket{0}_A\ket{0}_B\ket{0}_O +
\ket{1}_A\ket{1}_B\ket{1}_O)
$$
Thus, the oracle which implement a classical coin flipping protocol is
trivial regarding Definition~\ref{defn:trivial-oracle} and cannot help
to implement bit commitment.

Indeed, suppose that a quantum bit commitment protocol requires Alice
and Bob to share random coins at some steps. Recall
that just before the first call to the oracle, the quantum configuration
of the protocol, realized by two-party operations of Alice and
Bob, is in a state $\ket{\Psi(b)} = \sum_{i=1}^N \sqrt{p_b(i)}
\ket{i}_{E*}\ket{i}_A \ket{i}_B \ket{\psi_i(b)}_{AB}$,
cf. Eq. \eqref{equa:purified-classical-messages}. After receiving the
coins, the configuration becomes
$$
\ket{\Psi(b)}\otimes\ket{C} = \sum_{i=1..N,j=0..1} \sqrt{p_b(i)/2}
\ket{ij}_{E*}\ket{ij}_A \ket{ij}_B \ket{\psi_i(b)}_{AB}
$$
where $O$ is thrown to $E*$. Therefore, by induction, with any
successive unitary transformation on $A,B$ ands
request for random coins to the oracle, the global configuration of
the protocol remains in the penalized form,
cf. Eq. \eqref{equa:purified-classical-messages}. With this quantum
configuration, bit commitment and oblivious transfer are impossible.

From the view point of Copenhagen Interpretation as in Mayers'
proof~\cite{May97}, the quantum configuration of joint computation
just before a request to the coin flipping subroutine is a
projected state $\ket{\psi_i}_{AB}$ which is known to Alice and Bob
according to the exchanged messages $i$. Now, the coin flipping
subroutine provides either $\ket{0}_A\ket{0}_B$ or
$\ket{1}_A\ket{1}_B$ with equal probability. However, once the coins
are provided, Alice and Bob know which coin they have, and the global
state is accordingly a known state $\ket{\psi_i}_{AB} \otimes
\ket{0}_A\ket{0}_B$ or $\ket{\psi_i}_{AB} \otimes
\ket{1}_A\ket{1}_B$. And the no-go theorems can be applied to each of
these collapsed pure states.
\end{proof}

\section{Reversibility vs. Irreversibility}\label{sec:thermodynamics}
The topics of reversible computation are mostly studied in relation
with Landauer's principle of thermodynamical reversibility when resolving
the paradox of ``Maxell's demon:'' the erasure of one bit of
information in a computational device is necessarily accompanied by a
generation of $kT \ln 2$ heat~\cite{Lan61,Ben82,Bub01b,Ben03}.

\begin{figure}[htb]
\begin{center}
\resizebox{0.45\textwidth}{!}{\epsfig{file=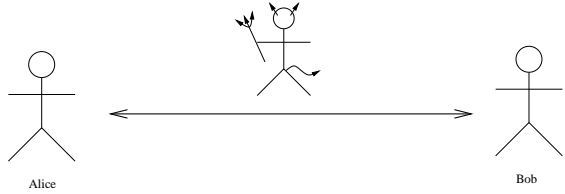}}
\caption{Secure two-party computations must be logically
information-erasing?}\label{fig:erasing-demon}
\end{center}
\end{figure}

A remarkable result from Theorems~\ref{thm:pe-oracle},
\ref{thm:trivial-oracle} is that,
unconditionally quantum secure oblivious transfer and bit commitment
can only
be made with help of a trusted third party which hide some
information from Alice and Bob. It implies that we have to have a
trusted third party which causes an logical erasure of information and
so, similar to Maxell's Demon, generates heat,
cf. Fig.~\ref{fig:erasing-demon}. It is
convenient to see that the third party has limited resource, and if
Alice and Bob invoke the request for many times, it begins to erase
its private memory by reset all to $\ket{0}$ or to overwrite its
memory and thus generate heat.

\begin{corollary}\label{corol:irreversibility}
Any quantum implementation of unconditionally secure oblivious transfer
and bit commitment requires erasure of information from the joint
views of Alice and Bob, and thus causes thermodynamical reversibility
and leads dissipation of heat to the environment.
\end{corollary}

One question is that: Is any process implementing unconditionally secure
oblivious transfer and bit commitment \emph{logically irreversible}?

An intuitive response from is Yes. Because, there are many positive
witnesses.

It was shown that any
logically reversible computation could be thermodynamically reversible
and implemented without heat dissipation, and vice versa, any
thermodynamically reversible computing process must be logically
reversible~\cite{Ben82,Ben00}. Moreover, it was shown that any
computation could be logically reversible, by Turing machine model
\cite{Ben73} or by logic circuit models~\cite{Tof80,FT82}. So,  all
two-party protocols are logically invertible:
\begin{itemize}
\item In a classical protocol, Alice and Bob can do any local computation
reversibly~\cite{Ben73}, for instance by using universal reversible
gates instead of normal irreversible gates AND, OR,
...~\cite{Tof80,FT82}. Therefore, the joint computation is a
reversible process over all variables at Alice and Bob locations.

\item In a quantum protocol, we expect that measurements will achieve
some erasure of information. However, Alice and Bob can keep all
of computations at the quantum level \emph{without measurement}, even
the final measurements because in an ideal protocol the users should
learn the results with certainty.
\item Then in the end of the protocols, Alice and Bob can make a copy
of the results, and undo all of the operations to reestablish the
thermodynamical condition.
\end{itemize}
This result is intuitively conformed to the impossibility of the
implementation of oblivious transfer and bit commitment by any
two-party protocol.

Evidently, when the users deny this behavior by throwing private
information then the erasure appears and we can build bit commitment
and  oblivious transfer protocols. For instance, we could implement
oblivious transfer by forcing Bob to measure the quantum
signals~\cite{Cre94,Yao95}. However, it is not that the erasure of
information is sufficient for implementing secure computations. For
instance, as analyzed in
Section~\ref{sec:macro-channel}, in a general two-party quantum
protocol with classical communication the measurements for making
classical messages can be logically seen as unnecessarily copying some
information to the external environment. The global process is then
\emph{logically reversible, though physically irreversible}.
In real protocols, we make unnecessary amplification of information to
the environment and cause unnecessary dissipation of heat.

Besides,  Rabin's oblivious transfer is equivalent to a
logical erasure channel. Thus implicitly, any logical process that
emulates Rabin OT would require the logical erasure of information,
such as noisy
channels~\cite{Cre97,CMW04}. And oblivious transfer may not be
implemented by any logically reversible computing process in the joint
view of Alice and Bob.

However, it's interesting to analyze this problem in two-party oracle
based protocols.

For quantum protocols using quantum oracles, the response is from
Corollary~\ref{corol:irreversibility}. We see that quantum two-party
oracle based protocols for oblivious transfer and bit commitment
require some entangled information, hidden or erased from the views of
Alice and Bob.

Nevertheless, we realize surprisingly that a classical oracle for
oblivious transfer, and so bit commitment, can be made with
unitary transitions. For instance, a simple classical circuit for
oblivious transfer with 2 input wires from Alice for $\{b_0,b_1\}_A$,
2 input wires from Bob for $\{c,x\}_B$, is built for the unitary
transition:
$$
\{b_0,b_1\}_A\{c,x\}_B \rightarrow \{b_0,b_1\}_1\{c, x\oplus b_c\}_B
$$
where $x$ is an auxiliary input for Bob to store the received
bit. This transition is one-to-one and so there exists a reverse
transition for it. Suppose that Alice and Bob send the inputs to the
oracle, get the outputs, make a copy of the result, and send the
outputs to an other oracle with the reverse transition which would
reestablish the thermodynamical condition for the first oracle.

So, could Alice and Bob realize oblivious transfer and bit commitment
for free, i.e. without dissipation of heat, by this way? Could
classical world beats the quantum one in this thermodynamical battle?

The response would be no, because the ultimate laws of macroscopic
behaviors are governed by quantum theory. Here, we must assume that
the classical oracle receives classical signals and treat them by a
unitary transformation. In other words, the classical oracle is
necessarily classical, acting in the classical world, not quantum
superposition one.

However, a process is necessarily classical only if it is collapsed to
the actual state of the environment. From this \emph{quantum view}, a
logical \emph{necessarily} classical bit is necessarily a observable
binary state, entangled with and amplified by the environment. This
observation leads some information to be stored somewhere in the outer
space, and must cause an entropy increase in the external environment.

\section{Conclusion}

In summary, we have proposed an detailed interpretation of general
quantum two-party protocols where the execution is seen as a
deterministic unitary evolution of a pure state covering all quantum
systems including Alice's and Bob's quantum dices purifying random
variables and local measurements, and environment's dices when a
macroscopic channel is used for transmitting classical information.

Thus, the global state is a pure three-party state, not two-party
state, where the environment's dices are not controllable by neither
Alice nor Bob. However, this impurity does not help to secure bit
commitment and oblivious transfer protocols. The state can be then
seen as a two-party one where the environment only amplifies classical
information given to the observer, while the other part can be fully
controlled by the cheater.
Therefore, the environment do not hide information from Bob in a
bit commitment protocol, and from Alice in an oblivious transfer
protocol. 

Obviously, secure two-party computations' primitives can be built with
help of trusted third-parties. However, we have shown that the no-go
theorems can also be applied to protocols that use trusted quantum
third-parties for computing any two-party function but which are
\emph{short-term}, i.e. they are built with limited resources and have
to throw information to the public environment; or \emph{post-empty},
i.e. they splits and redirects all output quantum variables to Alice
and Bob; or \emph{trivial}, i.e. they do computation with public
measurements only. Obviously, coin flipping belongs to this class of
trivial oracles.

These works implied that two-party oracles for implementing
unconditionally secure computations are required to
hide or erase information and considered as dissipation of heat.

\section{Acknowledgment}
We would like to thank Alain Maruani for helpful discussions.

\bibliographystyle{h-physrev}
\bibliography{quantum-biblio,crypto-biblio}

\end{document}

%% file: global-purified-model.pstex_t
\begin{picture}(0,0)%
\epsfig{file=global-purified-model.pstex}%
\end{picture}%
\setlength{\unitlength}{3947sp}%
\begingroup\makeatletter\ifx\SetFigFont\undefined%
\gdef\SetFigFont#1#2#3#4#5{%
  \reset@font\fontsize{#1}{#2pt}%
  \fontfamily{#3}\fontseries{#4}\fontshape{#5}%
  \selectfont}%
\fi\endgroup%
\begin{picture}(5764,3492)(3486,-5197)
\put(8712,-4170){\makebox(0,0)[lb]{\smash{{\SetFigFont{12}{14.4}{\familydefault}{\mddefault}{\updefault}{\color[rgb]{0,0,0}$D_B$}%
}}}}
\put(5480,-5197){\makebox(0,0)[lb]{\smash{{\SetFigFont{12}{14.4}{\familydefault}{\mddefault}{\updefault}{\color[rgb]{0,0,0}Classical channel's dices}%
}}}}
\put(7125,-1861){\makebox(0,0)[lb]{\smash{{\SetFigFont{12}{14.4}{\familydefault}{\mddefault}{\updefault}{\color[rgb]{0,0,0}$F_2$}%
}}}}
\put(5167,-1906){\makebox(0,0)[lb]{\smash{{\SetFigFont{12}{14.4}{\familydefault}{\mddefault}{\updefault}{\color[rgb]{0,0,0}$F_1$}%
}}}}
\put(4650,-3993){\makebox(0,0)[lb]{\smash{{\SetFigFont{12}{14.4}{\familydefault}{\mddefault}{\updefault}{\color[rgb]{0,0,0}A}%
}}}}
\put(7772,-3978){\makebox(0,0)[lb]{\smash{{\SetFigFont{12}{14.4}{\familydefault}{\mddefault}{\updefault}{\color[rgb]{0,0,0}B}%
}}}}
\put(5382,-3536){\makebox(0,0)[lb]{\smash{{\SetFigFont{12}{14.4}{\familydefault}{\mddefault}{\updefault}{\color[rgb]{0,0,0}$\ket{k}$}%
}}}}
\put(8069,-2941){\makebox(0,0)[lb]{\smash{{\SetFigFont{12}{14.4}{\familydefault}{\mddefault}{\updefault}{\color[rgb]{0,0,0}$\ket{j}$}%
}}}}
\put(4110,-3322){\makebox(0,0)[lb]{\smash{{\SetFigFont{12}{14.4}{\familydefault}{\mddefault}{\updefault}{\color[rgb]{0,0,0}$\ket{i}$}%
}}}}
\put(6255,-4251){\makebox(0,0)[lb]{\smash{{\SetFigFont{12}{14.4}{\familydefault}{\mddefault}{\updefault}{\color[rgb]{0,0,0}E*}%
}}}}
\put(3596,-4163){\makebox(0,0)[lb]{\smash{{\SetFigFont{12}{14.4}{\familydefault}{\mddefault}{\updefault}{\color[rgb]{0,0,0}$D_A$}%
}}}}
\end{picture}%